\documentclass[11pt]{article}
\pdfoutput=1
\usepackage{amsmath,amsfonts,amssymb,color}
\usepackage{cite,url}
\usepackage{graphicx}
\usepackage[utf8]{inputenc} 
\usepackage[bottom]{footmisc}

\allowdisplaybreaks

\def\nAk{-2}
\def\msklam{about $2.45$~TeV and $3$~TeV}
\def\msk2lam{$m_s\approx 5.5$~TeV, $m_a\approx 4.2$~TeV}

\def\smallSM{{\rm{\scriptscriptstyle SM}}}
\def\smallMSSM{{\rm{\scriptscriptstyle MSSM}}}
\def\smallNMSSM{{\rm{\scriptscriptstyle NMSSM}}}

\def\smallZ{{\scriptscriptstyle Z}}

\def\MS{M_S}
\def\MZ{m_\smallZ}

\def\beq{\begin{equation}}
\def\eeq{\end{equation}}
\def\bea{\begin{eqnarray}}
\def\eea{\end{eqnarray}}
\def\nn{\nonumber}
\def\wt{\widetilde}

\def\tbmssm{(\tan\beta)_\smallMSSM}

\def\lSM{\lambda_{\scriptscriptstyle{\rm SM}}}

\def\sq2{\sqrt{2}}
\def\drbar{{\ensuremath{ \overline{\rm DR}}}}
\def\msbar{\overline{\rm MS}}

\def\gl{\tilde{g}}
\def\mg{m_{\gl}}

\def\gs{g_s}

\long\def\symbolfootnote[#1]#2{\begingroup%
\def\thefootnote{\fnsymbol{footnote}}\footnote[#1]{#2}\endgroup}

\makeatletter
\newcommand{\vast}{\bBigg@{3}}
\makeatother

\begin{document}

\begin{titlepage}

\begin{flushright}
{\tt CERN-TH-2022-081}
\end{flushright}

\vspace{1cm}
\begin{center}

\vspace{1cm}

{\LARGE \bf Higgs-mass prediction in the NMSSM}\\[3mm]

{\LARGE \bf with heavy BSM particles} 

\vspace{1cm}

{\Large Emanuele Bagnaschi$^{\,a}$, Mark Goodsell$^{\,b}$ and
  Pietro~Slavich$^{\,b}$}

\vspace*{5mm}

{\sl ${}^a$ CERN, Theory Division, CH-1211 Geneva 23, Switzerland}
\vspace*{2mm}
\vspace*{2mm}\\{\sl ${}^b$
   Sorbonne Université, CNRS,
  Laboratoire de Physique Th\'eorique et Hautes Energies, 
 
  LPTHE, F-75005, Paris, France.}
\end{center}
\symbolfootnote[0]{{\tt e-mail:}}
\symbolfootnote[0]{{\tt Emanuele.Bagnaschi@cern.ch}}
\symbolfootnote[0]{{\tt goodsell@lpthe.jussieu.fr}}
\symbolfootnote[0]{{\tt slavich@lpthe.jussieu.fr}}

\vspace{0.7cm}

\abstract{We address the prediction for the mass of the SM-like Higgs
  boson in NMSSM scenarios where all BSM particles, including the
  singlets, have masses at the TeV scale. We provide a full one-loop
  computation of the matching condition for the quartic Higgs coupling
  in the NMSSM, supplemented with the two-loop contributions that
  involve the strong gauge coupling. We discuss the impact of the one-
  and two-loop corrections that are specific to the NMSSM on the
  prediction for the Higgs mass, and propose a method to estimate of
  the uncertainty associated with the uncomputed higher-order
  terms. Finally, we illustrate how the measured value of the Higgs
  mass can be used to constrain some yet-unmeasured parameters of the
  NMSSM.}

\vfill

\end{titlepage}

%\tableofcontents

\setcounter{footnote}{0}

\section{Introduction}
\label{sec:intro}

The discovery of a Higgs boson with mass around $125$~GeV and
properties compatible with the predictions of the Standard Model
(SM)~\cite{CMS:2012qbp, ATLAS:2012yve, ATLAS:2015yey, ATLAS:2016neq},
combined with the negative (so far) results of the searches for
additional new particles at the LHC, point to scenarios with at least
a mild hierarchy between the electroweak (EW) scale and the scale of
beyond-the-SM (BSM) physics. In this case, the SM plays the role of an
effective field theory (EFT) valid between the two scales. The
requirement that a given BSM model include a state that can be
identified with the observed Higgs boson can translate into important
constraints on the model's parameter space.

One of the prime candidates for BSM physics is supersymmetry (SUSY),
which predicts scalar partners for all SM fermions, as well as
fermionic partners for all bosons. A remarkable feature of SUSY
extensions of the SM is the requirement of an extended Higgs sector,
with additional neutral and charged bosons. In contrast to the case of
the SM, the masses of the Higgs bosons are not free parameters, as
SUSY requires all quartic scalar couplings to be related to the gauge
and Yukawa couplings. Moreover, radiative corrections to the
tree-level predictions for the quartic scalar couplings introduce a
dependence on all of the SUSY-particle masses and couplings. In a
hierarchical scenario such as the one described above, the lightest
scalar of the SUSY model plays the role of the SM Higgs boson, and the
prediction for its quartic self-coupling at the SUSY scale must
coincide with the SM coupling extracted at the EW scale from the
measured value of the Higgs mass and evolved up to the SUSY scale with
appropriate renormalization group equations (RGEs). This ``matching''
condition can be used to constrain some yet-unmeasured parameters of
the SUSY model, such as, e.g., the masses of the scalar partners of
the top quarks, the stops.

In the next-to-minimal SUSY extension of the SM, or NMSSM, the Higgs
sector includes two $SU(2)$ doublets $H_1$ and $H_2$ -- as in the case
of the minimal extension, the MSSM -- plus a complex scalar $S$,
singlet with respect to the SM gauge group.\footnote{For reviews of
  the MSSM and of the NMSSM, we point the reader to
  ref.~\cite{Martin:1997ns} and refs.~\cite{Maniatis:2009re,
    Ellwanger:2009dp}, respectively.}  The vacuum expectation value
(vev) of the singlet, induced by the mass and interaction terms in the
soft SUSY-breaking Lagrangian, generates a superpotential mass term
for the doublets. This provides a solution to the so-called ``$\mu$
problem'' of the MSSM, i.e., the question of why the supersymmetric
Higgs-mass parameter $\mu$ should be at the same scale as the soft
SUSY-breaking parameters. The doublet--singlet interactions of the
NMSSM also induce new contributions to the prediction for the quartic
self-coupling of the SM-like Higgs boson. Depending on the considered
region of the NMSSM parameter space, these additional contributions
can either increase or decrease the prediction for the quartic Higgs
coupling (and hence the Higgs mass) with respect to the case of the
MSSM.

The fixed-order (FO) calculation of the NMSSM Higgs boson masses -- in
which the Higgs self-energies are computed up to a given order in the
perturbative expansion, considering the full content of heavy and
light fields of the theory -- is quite advanced by now, albeit not yet
at the level of the corresponding calculation in the
MSSM.\footnote{For details on the Higgs-mass calculation in the MSSM,
as well as for an overview of the different approaches to the
Higgs-mass calculation in SUSY models, we point the reader to
ref.~\cite{Slavich:2020zjv}.} After early studies of the NMSSM Higgs
sector at the tree level~\cite{Ellis:1988er, Drees:1988fc} and partial
calculations of the dominant one-loop
corrections~\cite{Ellwanger:1993hn, Pandita:1993hx, Pandita:1993tg,
  Elliott:1993uc, Elliott:1993bs, King:1995vk, Ellwanger:2005fh,
  Ham:2001kf, Ham:2001wt, Ham:2003jf}, full calculations of the
one-loop corrections -- for increasingly general versions of the NMSSM
and using a variety of renormalization schemes for the NMSSM
parameters -- were made available in refs.~\cite{Degrassi:2009yq,
  Staub:2010ty, Ender:2011qh, Ross:2012nr, Graf:2012hh,
  Drechsel:2016jdg, Domingo:2017rhb, Hollik:2018yek, Dao:2021vqp}. At
the two-loop level, the corrections involving the strong gauge
coupling were computed in refs.~\cite{Degrassi:2009yq,
  Goodsell:2014bna, Goodsell:2015ira, Muhlleitner:2014vsa}, those
involving only the top Yukawa coupling were computed in
ref.~\cite{Dao:2019qaz}, and those involving also the remaining
superpotential couplings of the NMSSM were computed in
refs.~\cite{Goodsell:2014pla, Goodsell:2015ira, Goodsell:2016udb,
  Dao:2021khm}. It is worth pointing out that, in all of the two-loop
calculations listed above, the two-loop part of the Higgs
self-energies was computed under the approximation of vanishing
external momentum and in the so-called ``gaugeless limit'' of
vanishing EW gauge couplings. These full one-loop and partial two-loop
calculations of the NMSSM Higgs masses were implemented in various
public codes, such as
{\tt NMSSMTools}~\cite{Ellwanger:2004xm, Ellwanger:2005dv},
{\tt SARAH/SPheno}~\cite{Staub:2008uz, Staub:2009bi, Staub:2010jh,
  Staub:2012pb, Staub:2013tta,Porod:2003um, Porod:2011nf},
{\tt NMSSMCALC}~\cite{Baglio:2013iia},
{\tt SOFTSUSY}~\cite{Allanach:2001kg, Allanach:2013kza},
{\tt FlexibleSUSY}~\cite{Athron:2014yba, Athron:2017fvs} and
{\tt FeynHiggs}~\cite{Heinemeyer:1998yj, Hahn:2009zz,
  Bahl:2018qog}.\footnote{In fact, the version of {\tt FeynHiggs} that
implements the NMSSM calculations of refs.~\cite{Drechsel:2016jdg,
  Domingo:2017rhb, Hollik:2018yek} is not public yet.}
Detailed comparisons among the predictions of these codes for the
NMSSM Higgs masses were also presented in refs.~\cite{Staub:2015aea,
  Drechsel:2016htw}.

Compared with the case of the FO calculation, much less attention has
been devoted so far to the calculation of the NMSSM Higgs boson masses
in the EFT approach, in which the heavy fields are ``integrated out''
of the theory at a scale comparable to their mass, leaving behind
matching conditions for the couplings of the low-energy theory. This
approach is the most appropriate to scenarios characterized by a
hierarchy between the SUSY scale and the EW scale, where the FO
calculation is plagued by large logarithms of the ratio of the two
scales. In ref.~\cite{Athron:2016fuq} a method was proposed to
numerically obtain the boundary condition for the quartic Higgs
coupling at the SUSY scale, by matching the FO calculation of the pole
mass of the lightest Higgs boson in the NMSSM, as implemented in {\tt
  FlexibleSUSY}, with the corresponding calculation in the SM.  A
similar method was later implemented in {\tt SARAH/SPheno} in
ref.~\cite{Staub:2017jnp}.  The ``hybrid'' approach to the Higgs-mass
calculation in refs.~\cite{Athron:2016fuq, Staub:2017jnp} accounts for
terms suppressed by powers of $v^2/\MS^2$ -- where $v$ stands for the
EW scale and $\MS$ for the SUSY scale -- that would be neglected in a
pure EFT calculation. However, this approach required successive
adjustments~\cite{Athron:2017fvs, Kwasnitza:2020wli} to avoid the
introduction of spurious large-logarithmic effects at higher orders,
and it does not provide explicit analytic results for the matching
conditions. The latter would come in useful, e.g., to assess the
relevance of the various contributions, to check and compare existing
calculations, or as building blocks for further calculations.

A sensible approach to obtain explicit analytic results for the
matching conditions consists in adapting to the NMSSM the formulas
that were computed independently in refs.~\cite{Braathen:2018htl,
  Gabelmann:2018axh} for the one-loop matching of a general
high-energy theory (without heavy gauge bosons) on a general
renormalizable EFT. Indeed, in ref.~\cite{Gabelmann:2018axh} analytic
results for the one-loop matching condition for the quartic Higgs
coupling were obtained in an extremely constrained NMSSM scenario in
which all of the BSM-particle masses depend on just one
parameter. Subsequently, the general formulas of
ref.~\cite{Gabelmann:2018axh} were employed in
ref.~\cite{Gabelmann:2019jvz} to study a ``Split-SUSY'' scenario in
which the NMSSM is matched to an EFT that, beyond the SM fields,
includes also the complex singlet as well as the gauginos, the
higgsinos and the singlino (i.e., the SUSY partners of the gauge
bosons, the Higgs doublets and the singlet, respectively).

In this paper we aim to improve the accuracy of the EFT calculation of
the SM-like Higgs mass in the NMSSM, and to illustrate how the
measured value of the Higgs mass can constrain the parameter space of
the model even in scenarios where all of the BSM particles are
heavy. In section~\ref{sec:matching} we adapt to the NMSSM the general
formulas of ref.~\cite{Braathen:2018htl}, and obtain the full one-loop
matching condition for the quartic Higgs coupling, for arbitrary
values of all of the relevant parameters, in the EFT setup in which
the NMSSM is matched directly to the SM. We compare our results with
those of ref.~\cite{Gabelmann:2018axh}, in the constrained scenario
considered in that paper, and find a discrepancy. We also compute
directly the full two-loop-QCD contribution to the matching condition,
i.e., the contribution of all two-loop terms that involve the strong
gauge coupling. In section~\ref{sec:numbers} we discuss the effect of
the corrections computed in section~\ref{sec:matching} on the
prediction for the mass of the SM-like Higgs boson, including an
estimate of the uncertainty associated with uncomputed higher-order
terms. We also discuss the constraints on the NMSSM parameters that
arise from the requirement that the theory prediction for the Higgs
mass correspond to the measured value.  Finally,
section~\ref{sec:conclusions} contains our conclusions.

\section{Matching condition for the quartic Higgs coupling}
\label{sec:matching}
In this section we describe our calculation of the full one-loop and
two-loop-QCD contributions to the matching condition for the quartic
Higgs coupling in the NMSSM. 
To fix our notation, which follows the SLHA2
conventions~\cite{Allanach:2008qq}, we list here the terms in the
superpotential $W$ and in the soft SUSY-breaking Lagrangian ${\cal
  L}_{\rm soft}$ that determine the Higgs potential:
\bea
W&\supset& -\lambda\,\hat S\,\hat H_1\hat H_2\,+\, \frac{\kappa}3\,\hat S^3~,
\label{eq:Whiggs}\\
-{\cal L}_{\rm soft}
&\supset&
m_{H_1}^2 H_1^\dagger H_1\,+\,m_{H_2}^2 H_2^\dagger H_2\,+\,m_S^2\,S^* S
\,-\,\left(\lambda\,A_\lambda\,S\,H_1H_2\,-\, \frac{\kappa}3\,A_\kappa\,S^3
~+~{\rm h.c.}\right)~,
\label{eq:Lsoft}
\eea
where the hats denote superfields, and the $SU(2)$ indices of the
Higgs doublets are contracted by the antisymmetric tensor
$\epsilon_{ab}$, with $\epsilon_{12} =1$. For simplicity, we take the
parameters in eqs.~(\ref{eq:Whiggs}) and (\ref{eq:Lsoft}) to be all
real, and we enforce the $Z_3$ symmetry that forbids tadpole and mass
terms in $W$ and the corresponding SUSY-breaking terms in ${\cal
  L}_{\rm soft}$. We assume that the singlet develops a vev $v_s
\equiv \langle S \rangle$, providing effective $\mu$ and $B_\mu$ terms
for the Higgs doublets. It is also convenient to rotate the two
doublets to the so-called Higgs basis, in which they have the same
hypercharge and only one of them develops a vev
\beq
\left(\!\begin{array}{c} H \\ A \end{array}\!\right) =
\left(\!\begin{array}{rr} \cos
\beta&\!\sin\beta\\-\sin\beta
&\!\cos\beta \end{array}\!\right) \left(\!\!\!\begin{array}{c}
-\epsilon H_1^* \\~~\,H_2 \end{array}\!\right)~,
\eeq
where the rotation angle is defined by $\tan\beta\equiv v_2/v_1$, with
$v_i \equiv \langle H_i^0 \rangle$. In this basis the neutral component
of $H$ has the vev $v = (v_1^2 + v_2^2)^{1/2}$, while $A$ has no vev.

We consider the hierarchical scenario in which all of the SUSY
particles, as well as the Higgs doublet $A$ and the scalar and
pseudoscalar components of the singlet, are significantly heavier than
the EW scale. We then adopt an EFT setup in which all of the heavy
particles of the NMSSM are integrated out at a common renormalization
scale $Q\approx \MS$, below which the field content of the theory is
just the one of the SM, and in particular the Higgs doublet $H$ plays
the role of the SM Higgs. In the matching of the NMSSM to the SM we
work in the limit of unbroken EW symmetry, $v\rightarrow 0$. This
amounts to neglecting corrections suppressed by powers of $v^2/\MS^2$,
which can be mapped to the effect of non-renormalizable,
higher-dimensional operators in the EFT Lagrangian. In this limit we
can neglect the mixing among gauginos, higgsinos and 
singlino, as well as the mixing between the ``left'' and ``right''
sfermions (i.e., the SUSY partners of the left- and right-handed
fermions of the SM).

\subsection{Tree-level masses and couplings}
We now discuss the tree-level masses and couplings in the
Higgs/higgsino sector. As mentioned above, we work in the limit
$v\rightarrow 0$, as appropriate to the calculation of the
matching conditions in the EFT approach.
The masses of the scalar and pseudoscalar components of the singlet,
which we decompose as $S = v_s + (s + i\,a)/\sqrt 2$, are
\beq
\label{eq:msing}
m_s^2 ~=~\kappa \,v_s\,(A_\kappa + 4\,\kappa\,v_s)~,~~~~~~~
m_a^2 ~=~-3\,\kappa \,v_s\,A_\kappa~,
\eeq
where one of the minimum conditions of the tree-level scalar potential
has been used to replace the soft SUSY-breaking mass for the singlet
with a combination of the other parameters,
\beq
\label{eq:mintree}
m_S^2 ~=~ - \kappa\,v_s\,(A_k + 2\,\kappa\,v_s),
\eeq
and the requirement that $\langle S\rangle = v_s$ be a deeper minimum
than  $\langle S\rangle = 0$ implies
\mbox{$3\,\kappa\,v_s/A_\kappa  < - 1$}.
The mass of the heavy Higgs doublet $A$ is
\beq
\label{eq:mdoub}
m_A^2 = \frac{\lambda\,v_s\,(A_\lambda +
  \kappa\,v_s)}{\sin\beta\cos\beta}~,
\eeq
where again the minimum conditions have been used to remove the
dependence on $m^2_{H_1}$ and $m^2_{H_2}$. Finally, the higgsinos
$\tilde h_1$ and $\tilde h_2$ combine into a Dirac fermion with mass
$\mu = \lambda\, v_s$, and the singlino acquires a mass $m_{\tilde s}
= 2\,\kappa\,v_s$.

In the EFT approach, the calculation of the mass of the SM-like Higgs
boson $H$ -- or, alternatively, the determination of the constraints
on the NMSSM parameters that arise from the measured value of the
Higgs mass -- require the calculation of the NMSSM prediction for the
quartic Higgs coupling at the matching scale.\footnote{We denote the
quartic Higgs coupling as $\lSM$ to distinguish it from the
doublet--singlet superpotential coupling $\lambda$. In our conventions
the SM Lagrangian contains the quartic interaction term $-\frac12\lSM
|H|^4$\,.} At the tree level, the latter reads
\beq
\label{eq:lSMtree}
\lSM^{\scriptscriptstyle {\rm tree}} ~=~\frac{1}{4}(g^2 + g^{\prime\,2}) \cos^22\beta ~+~
\frac12\,\lambda^2 \sin^22\beta ~-~\frac{a_{hhs}^2}{m_s^2}~,
\eeq
where the first term on the r.h.s.~is the D-term contribution
analogous to the one in the MSSM, the second term is the F-term
contribution specific to the NMSSM, and the third term originates from
the decoupling of the singlet scalar. The trilinear coupling $a_{hhs}$
enters the NMSSM Lagrangian as ${\cal L} ~\supset -(a_{hhs}/2)\,h^2
s$, where $h$ is the neutral scalar component of $H$, and at the tree
level it reads
\beq
\label{eq:ahhs}
a_{hhs} ~=~ \frac{\lambda}{\sqrt2}\,
\left[2\,\lambda\,v_s - (A_\lambda + 2\,\kappa\,v_s)\,\sin2\beta\right]~.
\eeq
Combining eqs.~(\ref{eq:msing}), (\ref{eq:lSMtree}) and
(\ref{eq:ahhs}), we remark that the contribution to
$\lSM^{\scriptscriptstyle {\rm tree}}$ that arises from the decoupling
of the singlet is always negative, and contains a piece that does not
depend on $\tan\beta$.

\subsection{One-loop matching}
\label{sec:1loop}
We now describe our calculation of the full one-loop contribution to
the matching condition for the quartic Higgs coupling. This
contribution can be decomposed as:
\beq
\label{eq:dlam}
\Delta\lSM^{1\ell} ~=~ \Delta\lSM^{1\ell,{\scriptscriptstyle{\rm 1PI}}}
+\,2\,\lSM^{\scriptscriptstyle {\rm tree}}\,\Delta Z_h^{1\ell}
-\,2\,\frac{a_{hhs}}{m_s^2}\,\Delta a_{hhs}^{1\ell}
+\, \frac{a^2_{hhs}}{m_s^4}\, \Delta m_s^{2\,,\,1\ell}
+\,\Delta\lSM^{1\ell,{\scriptscriptstyle{\rm RS}}}
~.
\eeq
The first term on the r.h.s.~of eq.~(\ref{eq:dlam}) originates from
one-particle-irreducible (1PI) diagrams with heavy particles in the
loop and four external Higgs legs. The second term arises from
diagrams with a wave-function renormalization (WFR) insertion on one
of the external legs, and it involves the derivative with respect to
the external momentum of the heavy-particle (HP) contribution to the
renormalized self-energy of the Higgs field:
\beq
\label{eq:dZ}
\Delta Z_h^{1\ell} ~=~
-\left.\frac{d\hat\Pi_{hh}^{\scriptscriptstyle{\rm 1\ell,\,
      HP}}}{dp^2}\, \right|_{p^2=0}~.
\eeq
The third term on the r.h.s.~of eq.~(\ref{eq:dlam}) arises from
singlet-exchange diagrams with a HP loop insertion on one
of the trilinear Higgs--singlet vertices, while the fourth term arises
from singlet-exchange diagrams with a HP loop insertion on
the singlet propagator. The exact form of the latter term depends on
the choices made for the definitions of the singlet vev $v_s$ and of
the singlet mass $m_s^2$ entering the tree-level part of the matching
condition for $\lSM$, see eq.~(\ref{eq:lSMtree}). Assuming that $v_s$
corresponds to the minimum of the loop-corrected scalar potential, and
using the expression in eq.~(\ref{eq:msing}) for the singlet mass, we
get
\beq
\label{eq:dms}
\Delta m_s^{2\,,\,1\ell}~=~
-\left.\hat\Pi_{ss}^{\scriptscriptstyle{\rm 1\ell,\, HP}}
\right|_{p^2=0} +\, \frac{~~ \hat T_s^{\scriptscriptstyle{\rm 1\ell,\,
      HP}}}{\sqrt 2\,v_s}~,
\eeq
involving the HP contributions to the renormalized self-energy and
tadpole of the singlet. We remark that the tadpole term in
eq.~(\ref{eq:dms}) originates from the fact that, with our choice for
$v_s$, the minimum condition of the scalar potential used to remove
the explicit dependence of the singlet mass on the soft SUSY-breaking
parameter $m_S^2$ includes a loop correction,
\beq
\label{eq:minloop}
m_S^2 ~=~ - \kappa\,v_s\,(A_k + 2\,\kappa\,v_s) \,+\,
\frac{~~ \hat T_s^{\scriptscriptstyle{\rm 1\ell,\,
      HP}}}{\sqrt 2\,v_s}~.
\eeq
By expressing the singlet mass as in eq.~(\ref{eq:msing}), we are
effectively moving the tadpole term from the singlet mass to the
one-loop part of the matching condition for $\lSM\,$. If we were
instead to express the singlet mass entering the tree-level part of
the matching condition as $\,m_s^2 \,=\, m_S^2 \,+\, 2\, \kappa
\,v_s\,(A_\kappa + 3\,\kappa\,v_s)$, while still considering $v_s$ as
the minimum of the loop-corrected potential, there would be no tadpole
term in eq.~(\ref{eq:dms}). If, on the other hand, we were to consider
$v_s$ as the minimum of the tree-level potential, there would still be
no tadpole term in eq.~(\ref{eq:dms}), and the two expressions for the
singlet mass discussed above would be equivalent to each other,
i.e.~$m_s^2 = \kappa \,v_s\,(A_\kappa + 4\,\kappa\,v_s)= m_S^2 \,+\,
2\, \kappa \,v_s\,(A_\kappa + 3\,\kappa\,v_s)$. However, the one-loop
contribution to the matching condition for $\lSM$ in
eq.~(\ref{eq:dlam}) would receive direct contributions from non-1PI
diagrams with tadpole insertions (see ref.~\cite{Braathen:2021fyq} for
a related discussion)
\beq
\Delta\lSM^{1\ell,{\rm tad}}
~=~ 
\left[ \sqrt 2 \,\kappa\, (A_k + 6\,\kappa\,v_s)\,\frac{a_{hhs}^2}{m_s^6}
  - 2\,\lambda\,(\lambda -\kappa\,\sin 2\beta)\,\frac{a_{hhs}}{m_s^4}\right]
\,\hat T_s^{\scriptscriptstyle{\rm 1\ell,\, HP}}~,
\eeq
where the first term within square brackets arises from a diagram
with a tadpole insertion on the singlet propagator, and the second
from diagrams with tadpole insertions on the singlet-doublet vertices.

Finally, the last term on the r.h.s.~of eq.~(\ref{eq:dlam}) includes
contributions arising from differences in the renormalization scheme
(RS) used for the couplings of the NMSSM and for those of the EFT
valid below the matching scale (i.e., the SM). In particular, the
calculation of the matching condition for $\lSM$ is performed in the
$\drbar$ scheme assuming the field content of the NMSSM, whereas in
the EFT $\lSM$ is interpreted as an $\msbar$ parameter, and we also
choose to interpret the EW gauge couplings entering
eq.~(\ref{eq:lSMtree}) as $\msbar$ parameters of the SM. We remark,
however, that the presence of the singlet superfield does not affect
these contributions at one loop, thus they can be taken directly from
the MSSM calculation of refs.~\cite{Bagnaschi:2014rsa,
  Bagnaschi:2017xid}.

To obtain the quartic- and trilinear-vertex corrections and the
self-energies entering the various contributions to
eq.~(\ref{eq:dlam}) we use the formulas in appendix B of
ref.~\cite{Braathen:2018htl}, which discusses the one-loop matching of
a general high-energy theory (without heavy gauge bosons) on a general
renormalizable EFT. To obtain the singlet tadpole we use analogous
formulas from ref.~\cite{Braathen:2017izn}.  This saves us the trouble
of actually calculating one-loop Feynman diagrams, but requires that
we adapt to the case of the NMSSM the notation of
refs.~\cite{Braathen:2018htl,Braathen:2017izn} for masses and
interactions of scalars and fermions in a general renormalizable
theory.\footnote{Note that our conventions for the signs of tadpoles
and self-energies are opposite to those in
refs.~\cite{Braathen:2018htl,Braathen:2017izn}. Also,
refs.~\cite{Braathen:2018htl,Braathen:2017izn} define
$v_s\equiv\sqrt 2 \,\langle S\rangle$, whereas we define
$v_s\equiv\langle S\rangle$ as mentioned after eq.~(\ref{eq:Lsoft}).}
We find that, once the higgsino mass and the Higgs--sfermion
interaction parameters are expressed in terms of $\mu=\lambda\,v_s$,
the threshold correction $\Delta\lSM^{1\ell}$ splits neatly into a
part that does not depend explicitly on $\lambda$ and coincides with
the corresponding correction in the MSSM, see
refs.~\cite{Bagnaschi:2014rsa, Bagnaschi:2017xid}, plus an
NMSSM-specific, $\lambda$-dependent part that vanishes in the limit
$\lambda\rightarrow 0$.

Our full formulas for $\Delta\lSM^{1\ell}$, valid for generic values
of all of the relevant NMSSM parameters, are lengthy and not
particularly illuminating, therefore we make them available on request
in electronic form.  For later convenience, we provide here the
contribution to the NMSSM-specific part of $\Delta\lSM^{1\ell}$ from
diagrams involving stop squarks.  The sfermion contribution to
$\Delta\lSM^{1\ell,{\scriptscriptstyle{\rm 1PI}}}$ is the same as in
the MSSM once we set $\mu=\lambda\,v_s$, and in the limit
$v\rightarrow 0 $ there are no one-loop sfermion contributions to the
self-energy and tadpole of the singlet. The NMSSM-specific stop
contribution to the one-loop matching condition for $\lSM$ thus
reduces to
\beq
\label{eq:dl1lstop}
\left(\Delta\lSM^{1\ell,\,\tilde t}\,\right)_\lambda~=~
2\,(\lSM^{\scriptscriptstyle {\rm tree}})_\lambda \,\Delta Z_h^{1\ell,\,\tilde t}
-\,2\,\frac{a_{hhs}}{m_s^2}\,\Delta a_{hhs}^{1\ell,\,\tilde t}~.
\eeq
In the first term on the r.h.s.~of eq.~(\ref{eq:dl1lstop}) above,
$\left(\lSM^{\scriptscriptstyle {\rm tree}}\right)_\lambda = \lambda^2
\sin^22\beta/2 - a_{hhs}^2/m_s^2$ is the $\lambda$-dependent part of
the tree-level matching condition for $\lSM$, see
eq.~(\ref{eq:lSMtree}), and $\Delta Z_h^{1\ell,\,\tilde t}$ is the
stop contribution to the Higgs WFR:
\beq
\label{eq:dZ1l}
\Delta Z_h^{1\ell,\,\tilde t} ~=~ -\frac{g_t^2N_c}{(4\pi)^2}
\,\frac{X_t^2}{6\,m_{Q_3}m_{U_3}}\,\wt F_5\left(\frac{m_{Q_3}}{m_{U_3}}\right)~,
\eeq  
where $N_c=3$ is a color factor, $X_t = A_t - \mu\cot\beta$ is the
trilinear Higgs-stop interaction parameter (with $A_t$ being the
corresponding soft SUSY-breaking coupling), $m_{Q_3}$ and $m_{U_3}$
are the soft SUSY-breaking masses for the left and right stop,
respectively, and the loop function $\wt F_5(x)$ is defined in the
appendix A of ref.~\cite{Bagnaschi:2014rsa}.
In the second term on the r.h.s.~of eq.~(\ref{eq:dl1lstop}) above,
the tree-level quantities $m_s^2$ and $a_{hhs}$ are given in
eqs.~(\ref{eq:msing}) and (\ref{eq:ahhs}), respectively, while $\Delta
a_{hhs}^{1\ell,\,\tilde t}$ is the one-loop stop contribution to the
Higgs--singlet coupling in the limit $v\rightarrow 0$,
\beq
\label{eq:ahss1l}
\Delta a_{hhs}^{1\ell,\tilde t} ~=~
\sqrt2\,N_c\,\frac{\lambda\,g_t^2}{(4\pi)^2}\,
\frac{X_t\,\cot\beta}{m_{Q_3}^2-m_{U_3}^2}\,
\left[\,m_{Q_3}^2\left(1 - \ln\frac{m_{Q_3}^2}{Q^2}\right) -
m_{U_3}^2\left(1 - \ln\frac{m_{U_3}^2}{Q^2}\right)\right].
\eeq
The NMSSM-specific sbottom contribution
$\left(\Delta\lSM^{1\ell,\,\tilde b}\,\right)_\lambda$ can be obtained
from eqs.~(\ref{eq:dl1lstop})--(\ref{eq:ahss1l}) with the replacements
$g_t\rightarrow g_b$, $X_t\rightarrow X_b$,
$\cot\beta\rightarrow\tan\beta$ and $m_{U_3} \rightarrow m_{D_3} $,
where $g_b$ is the bottom Yukawa coupling, $X_b = A_b -
\mu\tan\beta\,$, and $m_{D_3}$ is the soft SUSY-breaking mass for the
right sbottom. Analogous replacements (with $N_c=1$) yield also the
stau contribution. For simplicity, we set the tiny Yukawa couplings of
the first two generations to zero, hence there are no NMSSM-specific
contributions to $\Delta\lSM^{1\ell}$ from the corresponding
sfermions.

As a non-trivial check of our full one-loop calculation, we verified
that by taking the derivative of the matching condition for $\lSM$
with respect to $\ln Q^2$ we recover the corresponding one-loop
RGE of the SM, i.e.,
\bea
(4\pi)^2\,\frac{d}{d\ln Q^2}\,
\left(\lSM^{\scriptscriptstyle {\rm tree}}+\Delta\lSM^{1\ell}\right) &=&
\lSM^{\scriptscriptstyle {\rm tree}}\left(6\,\lSM^{\scriptscriptstyle {\rm tree}}
\,+\, 6\,g_t^2\,+\,6\,g_b^2\,+\,2\,g_\tau^2\,-\,\frac92\,g^2
\,-\,\frac32 g^{\prime\,2}\right) \nn\\[1mm]
&& 
-\, 6\,g_t^4\,-\,6\,g_b^4\,-\,6\,g_\tau^4
\,+\,\frac{9}{8}\,g^4 \,+\, \frac{3}{8}\,g^{\prime\,4}
+ \frac34 \, g^2 g^{\prime\,2}~.
\eea
To this effect, we must combine the explicit scale dependence of our
result for $\Delta\lSM^{1\ell}$ with the implicit scale dependence of
all of the parameters entering $\lSM^{\scriptscriptstyle {\rm
    tree}}$. For the latter we use the RGEs of the NMSSM as listed in
ref.~\cite{Ellwanger:2009dp}, with the exception of the EW gauge
couplings for which our definitions require that we use the RGEs of
the SM.

As mentioned in section~\ref{sec:intro}, the authors of
ref.~\cite{Gabelmann:2018axh} obtained the matching condition for
$\lSM$ in an extremely constrained NMSSM scenario in which all of the
masses and couplings in the singlet/singlino sector depend only on
$\lambda$, $\tan\beta$ and a lone mass parameter $m_0$. After adapting
our formulas for $\Delta\lSM^{1\ell}$ to this constrained scenario, we
compared them with those of ref.~\cite{Gabelmann:2018axh},\footnote{We
  obtained the formulas of ref.~\cite{Gabelmann:2018axh} for
  $\Delta\lSM^{1\ell}$ in the constrained NMSSM scenario from their
  implementation in {\tt SARAH/SPheno}. We also remark that we cannot
  reproduce figure~12 of ref.~\cite{Gabelmann:2018axh} beyond the
  qualitative level.} but we found a discrepancy. Indeed, it appears
that in ref.~\cite{Gabelmann:2018axh} the one-loop matching condition
for $\lSM$ misses a tadpole contribution analogous to the one in our
eq.~(\ref{eq:dms}). This is inconsistent with the fact that the
authors of ref.~\cite{Gabelmann:2018axh} appear to have defined the
singlet mass entering $\lSM^{\scriptscriptstyle {\rm tree}}$ as in our
eq.~(\ref{eq:msing}) -- thus obtaining $m_s^2 = 2/3\,m_0^2\,$ in the
constrained scenario -- and $v_s$ as the vev of the loop-corrected
potential, as can be inferred from the absence of direct contributions
from tadpole-insertion diagrams in their results.

\subsection{Two-loop-QCD matching}
\label{sec:2loop}
As mentioned earlier, the one-loop squark contribution to the matching
condition for $\lSM$ splits into a $\lambda$-independent part that
coincides with the analogous MSSM result and a $\lambda$-dependent
part that is specific to the NMSSM. This structure allows for a
relatively economical calculation of the two-loop contribution that
involves the strong gauge coupling. Indeed, once we identify $\mu =
\lambda\,v_s$, the contribution from two-loop 1PI diagrams that
involve the strong interactions of the squarks is the same as in the
MSSM, and has already been computed in refs.~\cite{Bagnaschi:2014rsa,
  Bagnaschi:2019esc}. Furthermore, the fact that there are no squark
contributions to the singlet self-energy and tadpole at one loop
implies that there are no two-loop contributions involving the strong
interactions either. In analogy with eqs.~(\ref{eq:dlam}) and
(\ref{eq:dl1lstop}), the NMSSM-specific, $\lambda$-dependent
two-loop-QCD contribution to the matching condition for $\lSM$ can
thus be decomposed as
\beq
\label{eq:dlamtwo}
\left(\Delta\lSM^{2\ell,{\scriptscriptstyle{\rm QCD}}}\right)_\lambda
~=~
2\,\left(\lSM^{\scriptscriptstyle {\rm tree}}\right)_\lambda
\,\Delta Z_h^{2\ell,{\scriptscriptstyle{\rm QCD}}}
\,-~2\,\frac{a_{hhs}}{m_s^2}\,
\Delta a_{hhs}^{2\ell,{\scriptscriptstyle{\rm QCD}}}
\,+~\left(\Delta\lSM^{2\ell,{\scriptscriptstyle{\rm RS}}}\right)_\lambda~.
\eeq
The quantity denoted as $\Delta Z_h^{2\ell,{\scriptscriptstyle{\rm
      QCD}}}$ in the first term on the r.h.s.~of
eq.~(\ref{eq:dlamtwo}) above is the two-loop contribution to the Higgs
WFR from diagrams that involve the strong interactions of the
squarks. Again, once we identify $\mu = \lambda\,v_s$, this
contribution is the same as in the MSSM and has already been computed
in ref.~\cite{Bagnaschi:2019esc}. Explicit formulas for $\Delta
Z_h^{2\ell,{\scriptscriptstyle{\rm QCD}}}$ in the limit of degenerate
squark and gluino masses can be gleaned from eqs.~(13)--(17) in
section~2.2 of that paper, and the formulas with full dependence on
all of the relevant parameters are available on request in electronic
form. It is worth noting that, in the EFT calculation of the SM-like
Higgs mass, the WFR contribution in eq.~(\ref{eq:dlamtwo}) accounts
for effects that, in the corresponding FO calculation, arise from the
external-momentum dependence of the two-loop self-energies; in
particular, it is related to the first-order term in the
$p^2$-expansion of $\hat \Pi_{hh}^{2\ell, \tilde q}$, the contribution
to the renormalized Higgs-boson self-energy from two-loop diagrams
involving the strong interactions of the squarks. Higher-order terms
in the $p^2$-expansion of $\hat \Pi_{hh}^{2\ell, \tilde q}$ are
suppressed by powers of $v^2/\MS^2$ and can be neglected in our
hierarchic scenario, while the momentum dependence of the two-loop
quark--gluon contribution to the self-energy is fully accounted for in
the calculation of the relation between $\lSM$ and $M_h$ at the EW
scale, see section~\ref{sec:numbers}. Hence, the EFT approach allows
for a straightforward inclusion of external-momentum effects that are
of the same order in the relevant couplings as the other two-loop
contributions computed in this section. In contrast, these effects are
missed by the FO calculations of the NMSSM Higgs masses in
refs.~\cite{Degrassi:2009yq, Goodsell:2014bna, Goodsell:2015ira,
  Muhlleitner:2014vsa}, where the two-loop part of the self-energies
is computed -- whether ``diagrammatically'' or in the
effective-potential approach -- under the approximation of vanishing
external momentum.

The quantity denoted as $\Delta a_{hhs}^{2\ell,{\scriptscriptstyle{\rm
      QCD}}}$ in the second term on the r.h.s.~of
eq.~(\ref{eq:dlamtwo}) is the contribution to the trilinear
Higgs--singlet coupling from two-loop, 1PI diagrams that involve the
strong interactions of the squarks. This contribution is not available
in the literature, but can be straightforwardly computed with the same
effective-potential techniques adopted in
refs.~\cite{Bagnaschi:2014rsa, Bagnaschi:2019esc} for the calculation
of the two-loop-QCD contributions to the matching condition for
$\lSM$. In particular, we can write
\beq
\label{eq:ahssDV}
\Delta a_{hhs}^{2\ell,{\scriptscriptstyle{\rm QCD}}}
~=~ \left.
\frac{\partial^3 \Delta V^{2\ell,\,\tilde q}}{\partial^2 h\,\partial s }
\right|_{v=0}~,
\eeq
where $\Delta V^{2\ell,\,\tilde q}$ is the contribution to the NMSSM
scalar potential from two-loop diagrams involving the strong
interactions of the stop and sbottom squarks (the first two
generations of squarks do not contribute in the limit in which we
neglect the corresponding Yukawa couplings). The stop contribution to
$\Delta V^{2\ell,\,\tilde q}$ can be found, e.g., in eq.~(28) of
ref.~\cite{Bagnaschi:2014rsa}, and the sbottom contribution can be
obtained from the stop one with trivial replacements. The squark
masses and mixing angles in $\Delta V^{2\ell,\,\tilde q}$ are then
expressed as function of field-dependent quark masses, and
eq.~(\ref{eq:ahssDV}) becomes
\bea
\label{eq:ahssDV2}
\Delta a_{hhs}^{2\ell,{\scriptscriptstyle{\rm QCD}}}
&=&
-\,\frac{\lambda\,g_t^2}{\sqrt 2}\,\cot\beta\,\frac{d}{dX_t}
\left[\left( {\cal D}^{t\phantom b}_1 +\, m_t^2\,{\cal D}^t_2
  \right)\Delta V^{2\ell,\,\tilde t}\,\right]_{m_t\rightarrow 0}\nn\\[1mm]
&&-\,
\frac{\lambda\,g_b^2}{\sqrt 2}\,\tan\beta\,\frac{d}{dX_b}
\left[\left( {\cal D}^b_1 \,+\, m_b^2\,{\cal D}^b_2
  \right)\Delta V^{2\ell,\,\tilde b}\,\right]_{m_b\rightarrow 0}~,
\eea
where we define the operators
\beq
\label{eq:deriv}
      {\cal D}^q_i ~\equiv~\left( \frac{d}{dm_q^2}\right)^i.
\eeq      
We use relations such as those in eq.~(32) of
ref.~\cite{Bagnaschi:2014rsa} for the derivatives of the
field-dependent parameters with respect to the quark masses, then we
obtain the limits of vanishing quark masses as described in that paper
(note that in this calculation we do not encounter terms that diverge
for $m_q\rightarrow 0$). Finally, the derivatives with respect to the
parameters $X_t$ and $X_b$ in eq.~(\ref{eq:ahssDV2}) account for the
derivative with respect to the singlet field in
eq.~(\ref{eq:ahssDV}). In units of $g_s^2\,C_F\,N_c/(4\pi)^4$, where
$g_s$ is the strong gauge coupling and $C_F=4/3$ and $N_c=3$ are color
factors, we obtain for the stop contribution
\bea
\label{eq:ahssres}
\left(\Delta a_{hhs}^{2\ell,{\scriptscriptstyle{\rm QCD}}}\right)^{\tilde t}
&=&\!
- 2\sqrt2\,\lambda \,g_t^2\,\cot\beta\,\biggr\{\,
  X_t\left[
    -2 
    +\left(2 -\frac{2\,(1+x_Q)\ln x_Q}{x_Q-x_U}\right) \ln\frac{\mg^2}{Q^2}
    -\frac12 \ln^2\frac{\mg^2}{Q^2}
    \right.\nn\\[2mm]
    && ~~~~~~~~~~~~~~~~~~~~~~~~~~~~~
    +\frac{2 \,(1+2\,x_Q)\,\ln x_Q}{x_Q-x_U}
    -\frac{(x_Q^2 +x_Q-x_U)\,\ln^2 x_Q}{(x_Q-x_U)^2}\nn\\[2mm]
    && ~~~~~~~~~~~~~~~~~~~~~~~~~~~~~
    \left. + \frac{x_Q\,x_U\,\ln x_Q\,\ln x_U}{(x_Q-x_U)^2}
    - \frac{2\,(1-x_Q)}{x_Q-x_U}\,{\rm Li}_2\left(1-\frac1{x_Q}\right)\right]
  \nn\\[2mm]
  && ~~~~~~~~~~~~~~~~~~~+ \mg
  \left[ \,\frac52 
    -\left(2-\frac{2\,x_Q\,\ln x_Q}{x_Q-x_U}\right) \ln\frac{\mg^2}{Q^2}
    + \frac12 \ln^2\frac{\mg^2}{Q^2}-\frac{4\,x_Q\,\ln x_Q}{x_Q-x_U}
    \right.\nn\\[2mm]
    && ~~~~~~~~~~~~~~~~~~~~~~~~~~~~ \left.
    -\frac{(1-x_Q)\,\ln^2 x_Q}{x_Q-x_U} 
    - \frac{2\,(1-x_Q)}{x_Q-x_U}\,{\rm Li}_2\left(1-\frac1{x_Q}\right)\right]
  \nn\\[2mm]
  && ~~~~~~~~~~~~~~~~~~~+
  ~(x_Q \,\longleftrightarrow\, x_U)~~\biggr\}~,
\eea
where $\mg$ is the gluino mass, and we defined the ratios $x_Q =
m_{Q_3}^2/\mg^2\,$ and $x_U = m_{U_3}^2/\mg^2\,$. Again, the sbottom
contribution can be obtained from the stop contribution in
eq.~(\ref{eq:ahssres}) with the replacements $g_t\rightarrow g_b$,
$X_t\rightarrow X_b$, $\cot\beta\rightarrow\tan\beta$ and $m_{U_3}
\rightarrow m_{D_3} $.

The third term on the r.h.s.~of eq.~(\ref{eq:dlamtwo}) contains in
fact two separate contributions that arise from choices of
renormalization scheme for the couplings involved in the matching
condition:
\beq
\label{eq:scheme}
\left(\Delta\lSM^{2\ell,{\scriptscriptstyle{\rm RS}}}\right)_\lambda
~=~
\frac{g_s^2\,C_FN_c}{(4\pi)^4}\,(g_t^2+g_b^2)\,
\left(\lSM^{\scriptscriptstyle {\rm tree}}\right)_\lambda
\,+\,2\,\Delta g_t \left(\Delta\lSM^{1\ell,\,\tilde t}\,\right)_\lambda~.
\eeq
The first term on the r.h.s.~of eq.~(\ref{eq:scheme}) above is the
NMSSM-specific completion of the contribution in eq.~(21) of
ref.~\cite{Bagnaschi:2019esc}, and stems from the fact that SUSY
provides a prediction for the $\drbar$-renormalized quartic Higgs
coupling, whereas we interpret the parameter $\lSM$ in the EFT as
$\msbar$. The second term stems from the fact that we choose to
express the one-loop part of the matching condition for $\lSM$ in
terms of the $\msbar$-renormalized top Yukawa coupling of the SM
rather than the corresponding $\drbar$-renormalized coupling of the
NMSSM.\footnote{In contrast, the bottom Yukawa coupling is defined as
the $\drbar$-renormalized coupling of the NMSSM, to avoid the introduction of
spurious $\tan\beta$-enhanced effects at two loops (see
ref.~\cite{Bagnaschi:2017xid}).}
The quantity $\left(\Delta\lSM^{1\ell,\,\tilde
  t}\,\right)_\lambda$ is given in
eqs.~(\ref{eq:dl1lstop})--(\ref{eq:ahss1l}), while $\Delta g_t$
denotes the one-loop, ${\cal O}(g_s^2)$ part of the difference between
the $\msbar$ coupling of the SM and the $\drbar$ coupling of the
NMSSM,
\beq
\label{deltagt}
\Delta g_t ~=~ - \frac{\gs^2\,C_F}{(4\pi)^2}\,\left[1+
\ln\frac{\mg^2}{Q^2} 
\,+\, \wt F_6\left(\frac{m_{Q_3}}{\mg}\right)
\,+\, \wt F_6\left(\frac{m_{U_3}}{\mg}\right)
\,-\, \frac{X_t}{\mg}\,
\wt F_9\left(\frac{m_{Q_3}}{\mg},\frac{m_{U_3}}{\mg}\right)\right]~,
\eeq
where the loop functions $\wt F_6(x)$ and $\wt F_9(x,y)$ are defined
in the appendix A of ref.~\cite{Bagnaschi:2014rsa}.

Finally, we verified that, by taking the derivative with respect to
$\ln Q^2$ of the NMSSM-specific part of the two-loop-QCD matching
condition for $\lSM$, we do recover the terms involving $g_s^2$ that
we should expect from the RGE of the quartic Higgs coupling of the SM:
\beq
(4\pi)^4\,\frac{d}{d\ln Q^2}\,
\left[
  (\lSM^{\scriptscriptstyle {\rm tree}})_\lambda
  + \left(\Delta\lSM^{1\ell,\,\tilde q}\right)_\lambda
  + \left(\Delta\lSM^{2\ell,{\scriptscriptstyle{\rm QCD}}}\right)_\lambda
  \right]
~\supset~ 40\,g_s^2\,(g_t^2+g_b^2)\,(\lSM^{\scriptscriptstyle {\rm tree}})_\lambda~.
\eeq
To this effect, we must combine the explicit scale dependence of our
result for $\left(\Delta\lSM^{2\ell,{\scriptscriptstyle{\rm
      QCD}}}\right)_\lambda$ with the implicit scale dependence of the
parameters in $(\lSM^{\scriptscriptstyle {\rm tree}})_\lambda$ whose
RGEs have QCD contributions at two loops (namely $\lambda$,
$A_\lambda$ and $\tan\beta$), and of the parameters in the one-loop
stop and sbottom contributions $\left(\Delta\lSM^{1\ell,\,\tilde
  q}\right)_\lambda$ whose RGEs have QCD contributions at one loop
(namely the soft SUSY-breaking masses and trilinear couplings of the
squarks, and the top and bottom Yukawa couplings).

\vfill
\newpage
\section{Higgs-mass prediction and constraints on the NMSSM parameters}
\label{sec:numbers}

In this section we analyze the impact of the one- and two-loop
corrections to the quartic Higgs coupling computed in
sections~\ref{sec:1loop} and \ref{sec:2loop}, respectively, on the
prediction for the mass of the lightest Higgs scalar of the NMSSM,
focusing on the scenario in which all of the SUSY particles, the
heavier $SU(2)$ doublet, and the scalar and pseudoscalar components of
the singlet are all significantly heavier than the EW scale. We also
propose a method to estimate the so-called ``theory uncertainty'' of
our calculation, i.e., the uncertainty associated with uncomputed
higher-order corrections. Finally, we discuss how the prediction for
the mass of the SM-like Higgs boson can constrain the parameter space
of the NMSSM even in scenarios in which all of the BSM particles are
heavy.

\bigskip

Our numerical calculations rely on an EFT setup analogous to the one
developed in ref.~\cite{Bagnaschi:2019esc}. We use the public code
{\tt mr}~\cite{Kniehl:2016enc} to extract -- at full two-loop accuracy
-- the $\msbar$-renormalized parameters of the SM Lagrangian from a
set of physical observables, and to evolve them up to the SUSY scale
using the three-loop RGEs of the SM. For the physical observables
other than the Higgs mass, we use as input for the code the PDG values
$G_F= 1.1663787 \times 10^{-5}$~GeV$^{-2}$, $M_Z = 91.1876$~GeV, $M_W
= 80.379$~GeV, $M_t= 172.76$~GeV, $M_b = 4.78$~GeV and
$\alpha_s(\MZ)=0.1179$~\cite{ParticleDataGroup:2020ssz}.
In order to obtain a prediction for the Higgs mass from a full set of
SUSY parameters, we vary the value of the pole mass $M_h$ that we give
as input to {\tt mr} until the value of the $\msbar$-renormalized SM
parameter $\lSM(Q)$ returned by the code at the SUSY scale $Q=\MS$
coincides with the NMSSM prediction for the quartic coupling of the
lightest scalar. In alternative, we can treat the measured value of
the Higgs mass, $M_h = 125.25$~GeV~\cite{ParticleDataGroup:2020ssz},
as an additional input parameter, and use the matching condition on
the quartic Higgs coupling at the SUSY scale to constrain the NMSSM
parameter space. In this case we vary one of the SUSY parameters
%(e.g., a common mass term for the stops)
until the NMSSM prediction for the quartic Higgs coupling coincides
with the value of $\lSM(\MS)$ returned by {\tt mr}.

To obtain the NMSSM prediction for the quartic coupling of the SM-like
Higgs boson at the SUSY scale, we combine the tree-level prediction in
eq.~(\ref{eq:lSMtree}) with the full one-loop contribution computed in
section~\ref{sec:1loop}, the NMSSM-specific, $\lambda$-dependent
two-loop-QCD contribution computed in section~\ref{sec:2loop}, and the
$\lambda$-independent two-loop contributions that are in common with
the MSSM, given in refs.~\cite{Bagnaschi:2014rsa, Bagnaschi:2017xid,
  Bagnaschi:2019esc}. As a result, our determination of the matching
condition for $\lSM$ includes all of the two-loop contributions that
involve the strong gauge coupling, whereas the remaining two-loop
contributions are included only under the approximations of vanishing
EW gauge couplings and of vanishing $\lambda$. While the first
approximation is generally good in view of the relative sizes of $g$,
$g^\prime$, $g_t$ and $g_s$, the goodness of the second approximation
obviously depends on the considered values of $\lambda$. In particular
-- similarly to what was found in refs.~\cite{Goodsell:2014pla,
  Goodsell:2015ira, Goodsell:2016udb, Dao:2021khm} in the context of
the FO calculation of the NMSSM Higgs masses -- when $\lambda$ is of
${\cal O}(1)$ the two-loop, NMSSM-specific contributions that are
omitted in our prediction for $\lSM$ can become as large as the
dominant MSSM-like contributions that are included, i.e., those of
${\cal O}(g_t^4 g_s^2)$ and ${\cal O}(g_t^6)$. We recall that, in the
EFT approach, the (next-to)$^n$-leading-logarithmic (N$^n$LL)
resummation of the large logarithmic corrections requires the
combination of ($n\!+\!1$)-loop RGEs with $n$-loop matching conditions
at the boundary scales. Our calculation thus provides a full NLL
resummation of the large logarithmic corrections, but, in view of the
incomplete determination of the two-loop boundary condition for $\lSM$
at the SUSY scale, only a partial NNLL resummation. An estimate of the
impact of the omitted contributions seems desirable.

As was discussed already in ref.~\cite{Bagnaschi:2014rsa}, the EFT
calculation of the Higgs mass in scenarios where all heavy particles
are integrated out at a common scale $\MS$ is subject to three
distinct sources of theory uncertainty: $i)$ a ``SM uncertainty''
stemming from uncomputed higher-order corrections in the relations
between the physical observables taken as input for the calculation
and the $\msbar$-renormalized parameters of the SM Lagrangian; $ii)$ a
``SUSY uncertainty'' stemming from uncomputed higher-order corrections
in the matching conditions for the couplings of the SM Lagrangian at
the scale $\MS$; $iii)$ an ``EFT uncertainty'' associated to the
corrections suppressed by powers of $v^2/\MS^2$ that are omitted when
the EFT Lagrangian is identified with the renormalizable Lagrangian of
the SM in the unbroken phase of the EW symmetry. We neglect the third
source of uncertainty in the following, because we consider scenarios
in which all BSM particles are heavy enough to make the ${\cal
  O}(v^2/\MS^2)$ effects fully negligible. In contrast, we aim to
simulate the effects of uncomputed higher-order corrections in the
matching conditions at both the EW scale and the SUSY
scale.

To obtain an estimate of the SM uncertainty, we change the accuracy of
the determination of the top Yukawa coupling in {\tt mr}, removing
corrections of ${\cal O}(g_s^6)$ and higher that are implemented by
default in the code. This simulates the effect of uncomputed N$^3$LL
corrections that involve the highest powers of the strong gauge
coupling, which are expected to be the largest among those neglected
in the ``SM'' part of the calculation (see, e.g., the discussion in
section~6.3.1 of ref.~\cite{Slavich:2020zjv}).

For what concerns the SUSY uncertainty, we combine two estimates of
different classes of higher-order effects.  To obtain our first
estimate, which targets the uncomputed two-loop and higher-order
corrections that involve the top Yukawa coupling, we change the
definition of the coupling $g_t$ entering the matching condition for
$\lSM$ from the $\msbar$-renormalized parameter of the SM to the
$\drbar$-renormalized parameter of the NMSSM. The two definitions are
related by
\beq
\label{eq:shiftgt}
g_t^{\smallNMSSM}(\MS) ~=~ \frac{g_t^{\smallSM}(\MS)}{1- \Delta g_t^{\smallMSSM}
  - (\Delta g_t)_{\lambda}}~,
\eeq
where $\Delta g_t^{\smallMSSM}$ is the threshold correction given for
the MSSM in refs.~\cite{Bagnaschi:2014rsa, Bagnaschi:2017xid}, and
$(\Delta g_t)_{\lambda}$ is the $\lambda$-dependent contribution that
turns the MSSM coupling into the NMSSM coupling. This contribution is
related to the $\lambda$-dependent part of the Higgs WFR by $(\Delta
g_t)_{\lambda} = - (\Delta Z_h)_\lambda /2$. For the latter we find:
\bea
(4\pi)^2\,(\Delta Z_h)_\lambda &=&
-\frac{a_{hhs}^2}{2\,m_s^2}
\,-\,\frac{a_{hHs}^2}{6\,m_A\,m_s}\,\wt{F}_5\left(\frac{m_s}{m_A}\right)
\,-\,\frac{a_{hAa}^2}{6\,m_A\,m_a}\,\wt{F}_5\left(\frac{m_a}{m_A}\right)
\nn\\[2mm]
&& + \,\frac{\lambda^2}{3}\,\left[\,
  3\,\ln\frac{\mu^2}{Q^2}
  \,-\,\sin2\beta\,f\left(\frac{m_{\tilde s}}{\mu}\right)
  \,+\,g\left(\frac{m_{\tilde s}}{\mu}\right)\,\right]~,
\eea
where the terms in the first line are the contributions of diagrams
involving scalars or pseudoscalars, while those in the second line are
the contributions of diagrams involving higgsinos and singlino. The
loop functions $\wt F_5(x)$, $f(x)$ and $g(x)$ are defined in the
appendix A of ref.~\cite{Bagnaschi:2014rsa}. The coupling $a_{hhs}$ is
given in eq.~(\ref{eq:ahhs}), and the remaining trilinear couplings
are
\beq
a_{hHs} ~=~ -\frac{\lambda}{\sqrt 2}\,(A_\lambda + 2\,\kappa\,v_s)\,\cos 2\beta~,
~~~~~
a_{hAa} ~=~ \frac{\lambda}{\sqrt 2}\,(A_\lambda - 2\,\kappa\,v_s)~.
\eeq
The inclusion of this $\lambda$-dependent contribution in the
redefinition of $g_t$, see eq.~(\ref{eq:shiftgt}), simulates the
effect of two-loop contributions of ${\cal O}(g_t^4\lambda^2)$ to the
matching condition for $\lSM$. As mentioned above, these can in
principle compete with the known ${\cal O}(g_t^4 g_s^2)$ and ${\cal
  O}(g_t^6)$ contributions when $\lambda$ is large enough. We remark
that, to avoid the inclusion in our uncertainty estimate of effects
that are in fact accounted for by our calculation, the redefinition of
$g_t$ must be accompanied by shifts in the known two-loop
contributions to the matching condition for $\lSM$. In particular, in
the NMSSM-specific, $\lambda$-dependent two-loop-QCD contribution
computed in section~\ref{sec:2loop} we must remove the second term on
the r.h.s.~of eq.~(\ref{eq:scheme}). The analogous shifts for the
MSSM-like two-loop contributions can be trivially obtained from
refs.~\cite{Bagnaschi:2014rsa, Bagnaschi:2017xid, Bagnaschi:2019esc}.

Our second estimate of the SUSY uncertainty targets the uncomputed
two-loop and higher-order corrections that involve the highest powers
of the doublet--singlet coupling $\lambda$. We replace the definitions
in eq.~(\ref{eq:msing}) for the scalar and pseudoscalar singlet masses
with $\,m_s^2 \,=\, m_S^2 \,+\, 2\, \kappa \,v_s\,(A_\kappa +
3\,\kappa\,v_s)$ and $\,m_a^2 \,=\, m_S^2 \,+\, 2\, \kappa
\,v_s\,(-A_\kappa + \kappa\,v_s)$, respectively. The two sets of mass
definitions are equivalent at the tree level, but they differ at one
loop, because the soft SUSY-breaking singlet mass $m_S$ is related to
the other parameters as in eq.~(\ref{eq:minloop}) when $v_s$ is
defined as the vev of the loop-corrected potential. As discussed in
section~\ref{sec:1loop}, the change in the definition of the scalar
singlet mass entering the tree-level part of the matching condition
for $\lSM$ must be compensated for at the one-loop level by removing
the tadpole term in eq.~(\ref{eq:dms}). On the other hand, the change
in the definition of both scalar and pseudoscalar singlet masses
entering the one-loop part of the matching condition simulates the
effect of two-loop $\lambda$-dependent corrections that do not involve
the strong gauge coupling, and the change in the definition of the
scalar singlet mass entering the two-loop-QCD part simulates the effect of
three-loop $\lambda$-dependent corrections that do involve the strong
gauge coupling.

\bigskip

\begin{figure}[t]
\begin{center}
  \vspace*{-1.2cm}
  \includegraphics[width=8.3cm]{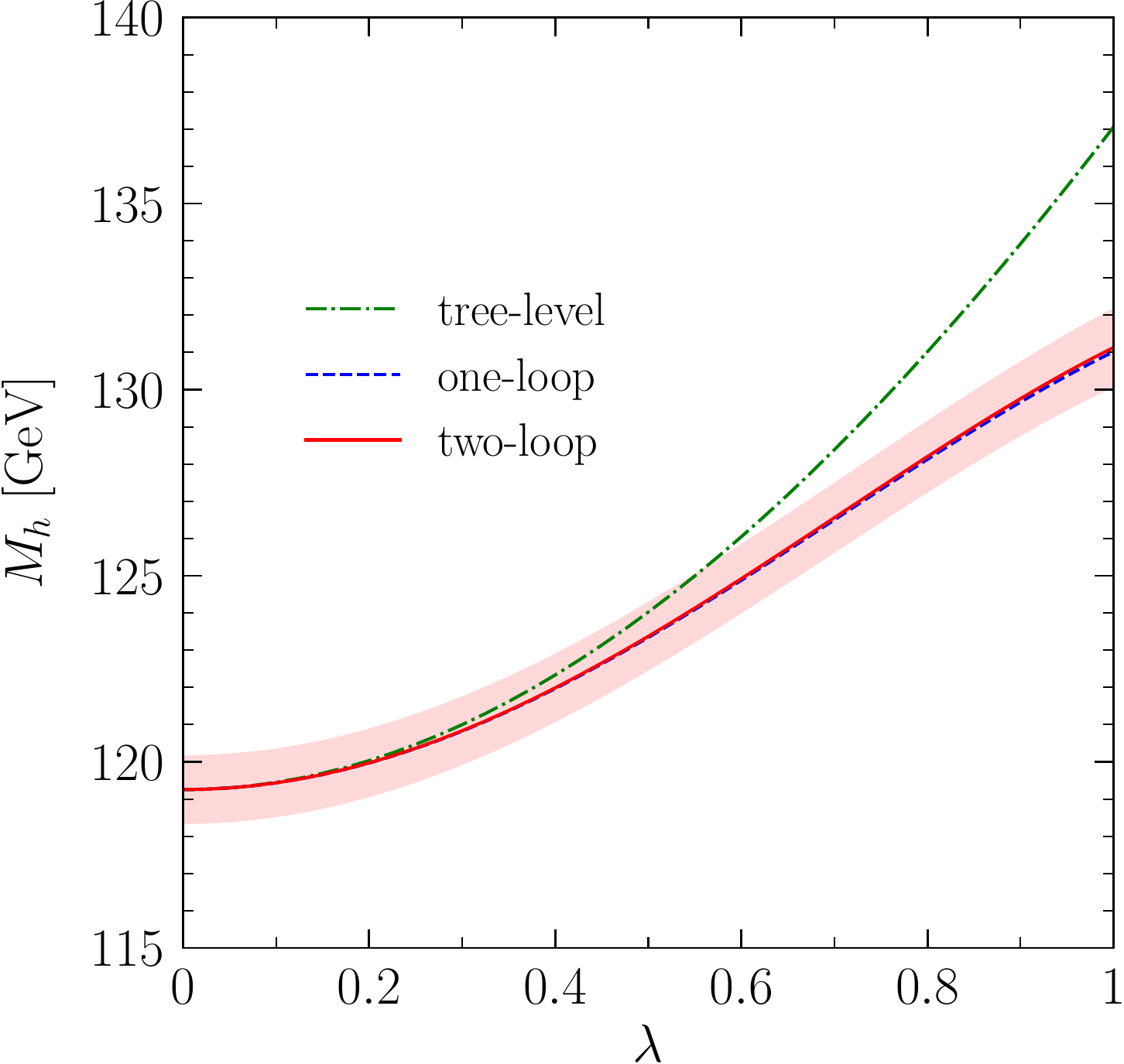}~~~
  \includegraphics[width=8.3cm]{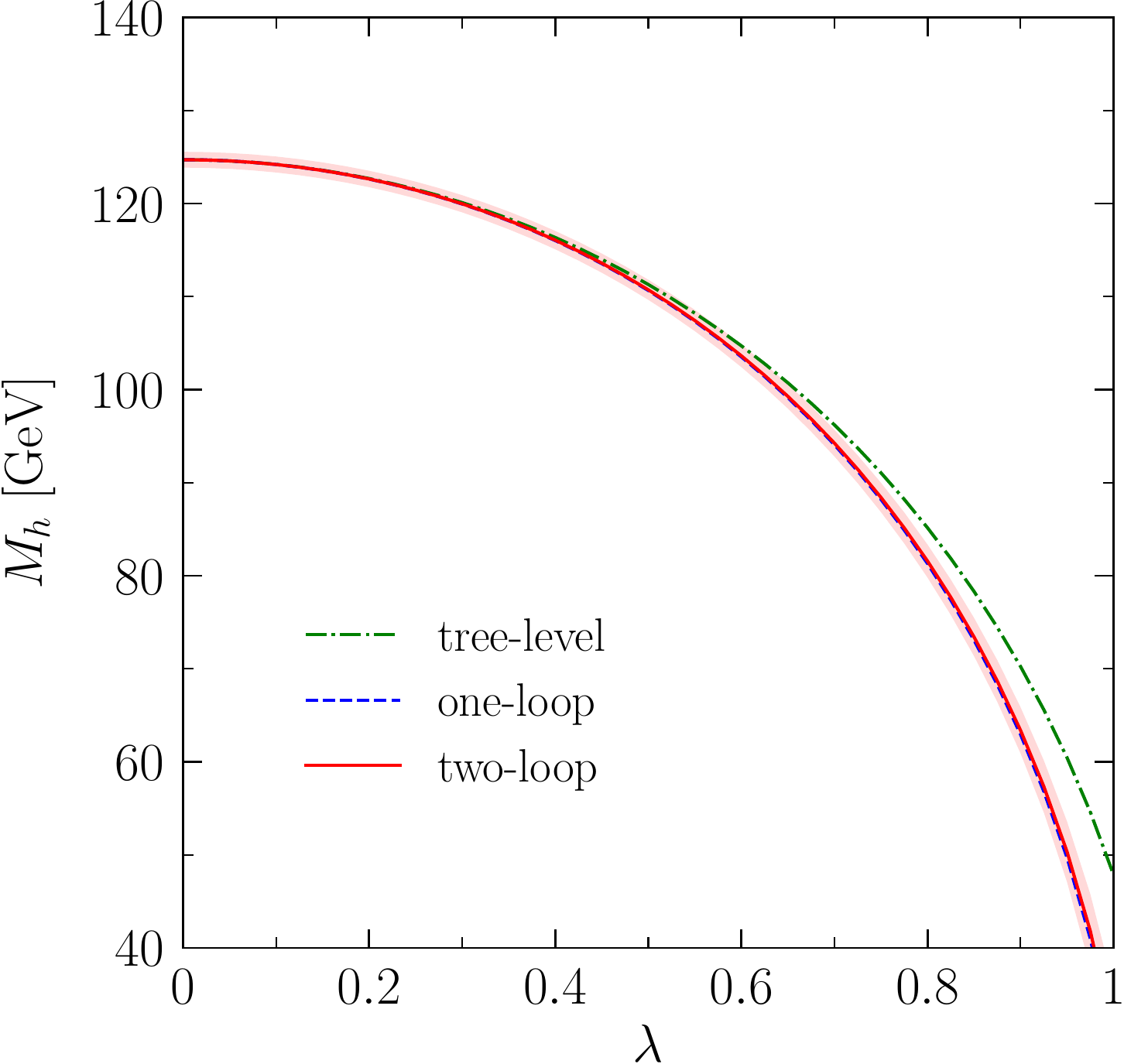}
  \caption{\em Higgs-mass prediction as a function of $\lambda$, for
    $\tan\beta=3$ (left) or $\tan\beta=5$ (right), in the NMSSM
    scenario described in the text. The three lines in each plot
    correspond to different accuracies of the NMSSM-specific
    contribution to the matching condition for $\lSM$. The band around
    the red, solid line is our estimate of the theory
    uncertainty.}
  \label{fig:mhvslam}
\vspace*{-5mm}
\end{center}
\end{figure}

To assess the impact of the different contributions to the matching
condition for the quartic Higgs coupling computed in
section~\ref{sec:matching}, we plot in figure~\ref{fig:mhvslam} the
prediction for the mass of the SM-like Higgs boson as a function of
$\lambda$. We consider an NMSSM scenario in which all sfermions of the
first and second generations have degenerate masses of $2$~TeV, while
those of the third generation (namely stops, sbottoms and staus) have
degenerate masses $\MS = 5$~TeV. The trilinear Higgs-stop interaction
parameter is fixed as $X_t = \sqrt6\,\MS$\,, which maximizes the
one-loop stop contribution to $\lSM$. For given values of $\mu$ and
$\tan\beta$ this determines the soft SUSY-breaking coupling $A_t$, and
the corresponding sbottom and stau couplings are fixed as
$A_b=A_\tau=A_t$. For consistency with our calculation of the two-loop
contributions to $\lSM$, all sfermion masses and trilinear couplings
are interpreted as \drbar-renormalized parameters, at a scale that we
choose as $Q=\MS$. The soft SUSY-breaking masses of bino, wino and
gluino are fixed as $M_1=1$~TeV, $M_2=2$~TeV and $M_3=2.5$~TeV,
respectively.

In contrast to the case of the MSSM, in which the tree-level masses of
the heavy Higgs bosons and of the higgsinos in the limit of unbroken
EW symmetry are determined by the three parameters $\mu$, $B_\mu$ and
$\tan\beta$ (with $m_A^2 = 2\,B_\mu/\sin2\beta$), the Higgs/higgsino
sector of the NMSSM depends on the six parameters $\lambda$, $\kappa$,
$v_s$, $A_\lambda$, $A_\kappa$ and $\tan\beta$. In the plots of
figure~\ref{fig:mhvslam} we choose to vary the doublet--singlet
coupling $\lambda$, because that parameter determines the size of the
NMSSM-specific contributions to the quartic Higgs coupling, and in the
limit $\lambda\rightarrow 0$ we recover the MSSM prediction. For the
remaining parameters, we choose: $\kappa = \lambda$; a tree-level
higgsino mass $\mu = \lambda\,v_s = 1.5$~TeV, which determines $v_s$
for a given value of $\lambda$; a tree-level heavy-Higgs-doublet mass
$m_A = 3$~TeV, which determines $A_\lambda$ via eq.~(\ref{eq:mdoub})
for a given value of $\tan\beta$; and $A_\kappa = \nAk$~TeV. Finally,
we fix $\tan\beta=3$ in the left plot of figure~\ref{fig:mhvslam} and
$\tan\beta=5$ in the right plot. For consistency with our calculation
of the one-loop contributions to $\lSM$, all of these six parameters
-- which enter the boundary condition for $\lSM$ already at the tree
level -- are interpreted as \drbar-renormalized parameters, also at
the scale $Q=\MS$. Our choices of parameters correspond to a
tree-level mass of $3$~TeV for the singlino, and to tree-level masses
of \msklam\ for the scalar and pseudoscalar components of the singlet,
respectively.

In all of the lines in the plots of figure~\ref{fig:mhvslam}, the
Higgs-mass prediction includes all of the contributions to the
matching condition for $\lSM$ that are in common with the MSSM, so
that the left edge of the plot where $\lambda = 0$ corresponds to our
best prediction for $M_h$ in the so-called ``MSSM limit''. The three
lines in each plot correspond to different accuracies for the
inclusion of the NMSSM-specific contributions. The green, dot-dashed
line corresponds to the inclusion of the tree-level contribution
$\left(\lSM^{\scriptscriptstyle {\rm tree}}\right)_\lambda = \lambda^2
\sin^22\beta/2 - a_{hhs}^2/m_s^2\,$ alone; the blue, dashed line
corresponds to the inclusion of the full one-loop, $\lambda$-dependent
contribution computed in section~\ref{sec:1loop}; the red, solid line
corresponds to the additional inclusion of the two-loop-QCD,
$\lambda$-dependent contribution computed in section~\ref{sec:1loop}.
The band around the red, solid line corresponds to our total estimate
of the theory uncertainty of the Higgs-mass prediction, obtained by
summing linearly the absolute values of the three estimates described
above. Comparing the different contributions, we find that the SM
uncertainty estimate alone is generally larger than the combination of
the two SUSY uncertainty estimates.

The comparison between the left and right plots in
figure~\ref{fig:mhvslam} shows that, in our scenario, the
$\lambda$-dependent contributions increase the prediction for $M_h$
for lower values of $\tan\beta$, and decrease it for higher values of
$\tan\beta$. This behavior is driven already at the tree level by the
$\tan\beta$ dependence of $\left(\lSM^{\scriptscriptstyle {\rm
    tree}}\right)_\lambda$, whose first, positive-definite term is
suppressed at larger $\tan\beta$, whereas the second, negative-definite
term contains a $\tan\beta$-independent piece, as remarked after
eq.~(\ref{eq:ahhs}). The comparison between the dot-dashed and dashed
lines in each plot shows that the full one-loop, $\lambda$-dependent
contribution to the matching condition for $\lSM$ can become
substantial when $\lambda \gtrsim 0.5$, changing the prediction for
$M_h$ by several GeV. Finally, the comparison between the dashed and
solid lines shows that the effect on the Higgs-mass prediction of the
two-loop-QCD, $\lambda$-dependent contribution is quite modest, and it
is much smaller than our estimate of the uncomputed higher-order
effects. This is likely related to the fact that, with our choices of
parameters, the $\lambda$-dependent stop contribution is suppressed
already at the one-loop level. In particular, the second term in
eq.~(\ref{eq:dl1lstop}) vanishes for degenerate stop masses and
$Q=\MS$, and the first term only amounts to a $2\%$ shift of the
tree-level contribution.

\bigskip

\begin{figure}[t]
\begin{center}
  \vspace*{-1.2cm}
  \includegraphics[width=8.2cm]{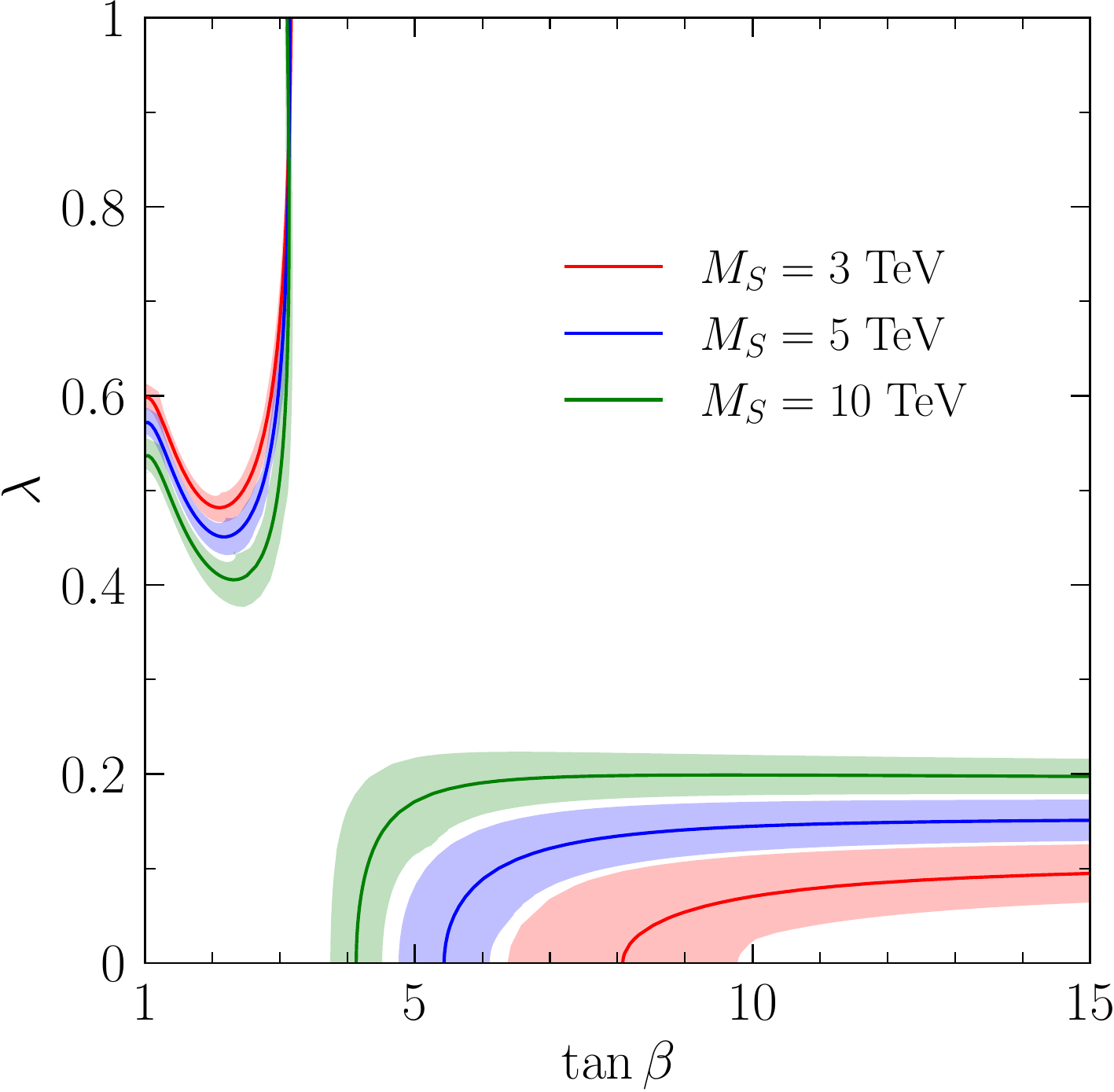}~~~
  \includegraphics[width=8.2cm]{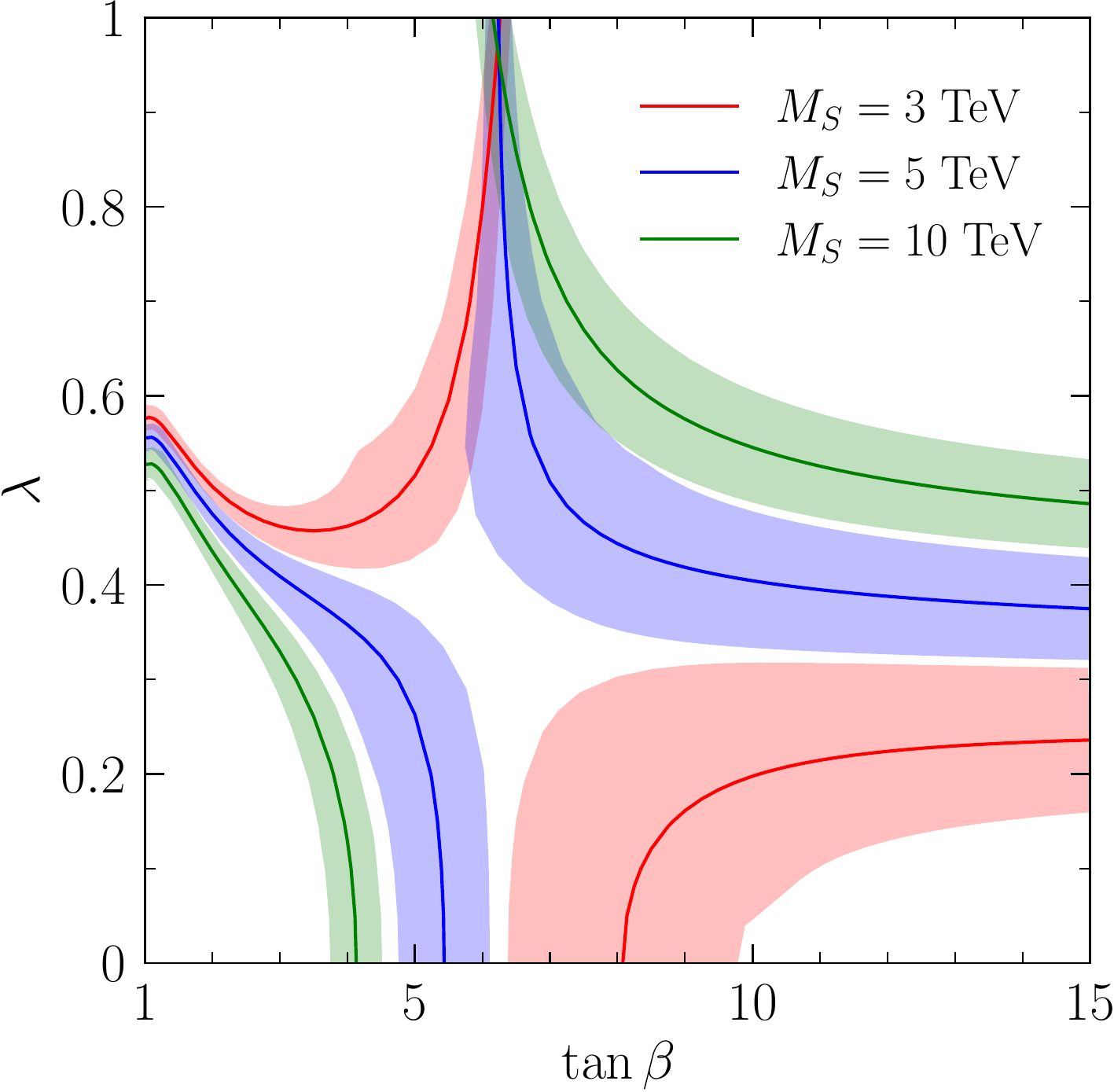}
  \caption{\em Regions of the $\tan\beta$--$\lambda$ plane that yield
    $M_h=125.25$~GeV, for $\kappa=\lambda$ (left) or
    $\kappa=2\,\lambda$ (right), in the NMSSM scenario described in
    the text. The three sets of lines in each plot correspond to
    different values of the common mass term for the third-generation
    sfermions. The band around each line is our estimate of the theory
    uncertainty.}
  \label{fig:scanlamtb}
\vspace*{-5mm}
\end{center}
\end{figure}

The knowledge of the mass of the SM-like Higgs boson can be used to
constrain the parameters of the yet-undiscovered SUSY sector. For
example, figure~3 of ref.~\cite{Bagnaschi:2019esc} showed the values
of $\MS$ and $X_t$ -- i.e., the two parameters that most affect the
stop contribution to the matching condition for $\lSM$ -- that are
selected by the requirement that the prediction for $M_h$ coincide
with its measured value in an MSSM scenario with moderately large
$\tan\beta$. Here we keep the parameters in the stop sector fixed, and
exploit the Higgs-mass prediction to constrain two of the parameters
that determine the NMSSM-specific contribution to the matching
condition for $\lSM$ already at the tree level. In
figure~\ref{fig:scanlamtb} we show the values of $\tan\beta$ and
$\lambda$ that yield the required prediction $M_h=125.25$~GeV in a
representative NMSSM scenario with heavy BSM particles. We adopt the
same choices of SUSY parameters as in figure~\ref{fig:mhvslam}, except
that $i)$ we set $\MS$ to $3$~TeV (red lines), $5$~TeV (blue lines) or
$10$~TeV (green lines), and $ii)$ we set either $\kappa=\lambda$ (left
plot) or $\kappa = 2\,\lambda$ (right plot).\footnote{For $\kappa =
  2\,\lambda$ the tree-level masses of the singlet fields become
  \msk2lam\ and $m_{\tilde s} = 6$~TeV.}  Each of the lines in
figure~\ref{fig:scanlamtb} is obtained with our full one-loop and
partial two-loop calculation of the matching condition for $\lSM$, and
is accompanied by an uncertainty estimate obtained as described
earlier.\footnote{The kinks in figure~\ref{fig:scanlamtb} are
  artifacts induced by the symmetrization of the uncertainty bands in
  the $\tan\beta$--$\lambda$ plane.}

The behavior of the different lines in the plots of
figure~\ref{fig:scanlamtb} can be qualitatively understood by
considering the dependence on $\tan\beta$ and $\lambda$ of the three
terms entering the tree-level matching condition for $\lSM$, see
eq.~(\ref{eq:lSMtree}). The first term, which is in common with the
MSSM, vanishes for $\tan\beta=1$ and increases for increasing
$\tan\beta$, reaching a constant positive value at large
$\tan\beta$. With our choices for the parameters that determine
$a_{hhs}$ and $m_s$, the remaining, NMSSM-specific contribution
$\left(\lSM^{\scriptscriptstyle {\rm tree}}\right)_\lambda = \lambda^2
\sin^22\beta/2 - a_{hhs}^2/m_s^2\,$ scales as $\lambda^2$, with a
coefficient that is positive at low values of $\tan\beta$, decreases
for increasing $\tan\beta$, eventually turns negative, and finally
reaches a constant negative value at large $\tan\beta$.  In each of
the plots of figure~\ref{fig:scanlamtb}, the line corresponding to a
given value of $\MS$ is split into a left branch at lower $\tan\beta$,
where the NMSSM-specific contribution to the matching condition for
$\lSM$ is positive, and a right branch at higher $\tan\beta$, where
the NMSSM-specific contribution is negative. The point where each line
meets the $x$ axis corresponds to a value of $\tan\beta$ that we
denote as $\tbmssm$, for which the required prediction for $M_h$ is
obtained in the ``MSSM limit'' $\lambda \rightarrow 0$. This depends
on the value of $\MS$, because heavier stops induce a larger positive
contribution to the matching condition for $\lSM$ and thus allow for
lower $\tan\beta$ in the MSSM limit.

In the left plot of figure~\ref{fig:scanlamtb}, where we set
$\kappa=\lambda$, the NMSSM-specific contribution to the matching
condition for $\lSM$ turns negative for a value of $\tan\beta$ that is
lower than $\tbmssm$. At the left edge of the plot, where
$\tan\beta=1$, the MSSM prediction for the Higgs mass is too low, and
the required prediction $M_h=125.25$~GeV is obtained thanks to the
positive NMSSM-specific contribution. Moving to higher values of
$\tan\beta$, the value of $\lambda$ that yields the required
prediction for $M_h$ decreases at first, as the tree-level, MSSM-like
contribution to $\lSM$ increases. However, the coefficient of
$\lambda^2$ in the tree-level,\footnote{We refer only to the
tree-level contribution for a qualitative interpretation of the
plots. The presence of the radiative corrections, some of which scale
as $\lambda^4$ in our scenario, does not alter the overall behavior
described here.}  NMSSM-specific contribution decreases for increasing
$\tan\beta$, and as it approaches zero the value of $\lambda$ that
yields $M_h=125.25$~GeV shoots up.  In the gap between the value of
$\tan\beta$ for which the NMSSM-specific contribution turns negative
and $\tbmssm$ there is no value of $\lambda$ that yields the required
Higgs-mass prediction. Finally, when $\tan\beta$ is larger than
$\tbmssm$ the MSSM prediction for $M_h$ is too high, and the required
prediction is obtained thanks to the negative NMSSM-specific
contribution. As both the coefficient of $\lambda^2$ in
$\left(\lSM^{\scriptscriptstyle {\rm tree}}\right)_\lambda$ and the
MSSM prediction for $M_h$ reach a plateau at large $\tan\beta$, so
does the value of $\lambda$ that brings the Higgs mass down to
$125.25$~GeV.

In the right plot of figure~\ref{fig:scanlamtb}, where we set
$\kappa=2\,\lambda$, the qualitative behavior of the red line
corresponding to $\MS=3$~TeV is the same as in the left plot. However,
the blue and green lines corresponding to higher values of $\MS$
behave differently. In this case, the value of $\tan\beta$ for which
the NMSSM-specific contribution to the matching condition for $\lSM$
turns negative is higher than $\tbmssm$. Consequently, as $\tan\beta$
approaches $\tbmssm$ from the left, the requirement that
$M_h=125.25$~GeV drives $\lambda$ to zero. On the right of $\tbmssm$
there is a gap in which the MSSM prediction for $M_h$ is too high but
the NMSSM-specific contribution remains positive, so there is no value
of $\lambda$ that yields the required Higgs-mass prediction. Finally,
when the NMSSM-specific contribution does turn negative, the value of
$\lambda$ that brings the prediction for $M_h$ down to $125.25$~GeV
decreases with increasing $\tan\beta$, and eventually reaches a
plateau as in the left plot.

\section{Conclusions}
\label{sec:conclusions}

If SUSY is realized in nature, there appears to be at least a mild
hierarchy between the scale of the superparticle masses and the EW
scale. In this kind of hierarchical scenario the prediction for the
mass of the SM-like Higgs boson is best obtained in the EFT approach,
which allows for the all-orders resummation of potentially large
corrections enhanced by powers of the logarithm of the two scales. The
EFT calculation of the Higgs masses in the MSSM is by now quite
advanced, with full one-loop and partial two-loop results for the
matching conditions for the Higgs self-couplings under a variety of
mass hierarchies (see ref.~\cite{Slavich:2020zjv} for a review). In
contrast, in the case of the NMSSM analytic calculations of the
matching conditions have been performed so far only at the one-loop
level, in an extremely constrained scenario where all BSM particles
are heavy and their masses depend on just one
parameter~\cite{Gabelmann:2018axh}, and in a Split-SUSY scenario where
the low-energy EFT includes also the scalar and pseudoscalar components
of the singlet plus all of the SUSY fermions~\cite{Gabelmann:2019jvz}.

In this paper we obtained a full one-loop result, valid for arbitrary
values of all relevant parameters, for the matching condition for the
quartic coupling of the Higgs boson, in the NMSSM scenario where all
BSM particles are heavy and the EFT valid below the SUSY scale is just
the SM. To this purpose, we adapted to the NMSSM the results of
ref.~\cite{Braathen:2018htl}, which provides the one-loop matching of
a general high-energy theory (without heavy gauge bosons) on a general
renormalizable EFT. We compared our results with those of
ref.~\cite{Gabelmann:2018axh} -- in the constrained scenario discussed
in that paper -- and found a discrepancy related to the definition of
the singlet mass entering the tree-level part of the matching
condition for $\lSM$. In addition to the full one-loop calculation of
the matching condition, we directly computed the two-loop
contributions that involve the strong gauge coupling. Our result
includes also terms associated with the external-momentum dependence
of the two-loop Higgs self-energy that are missed by the FO
calculations of the corresponding corrections in
refs.~\cite{Degrassi:2009yq, Goodsell:2014bna, Goodsell:2015ira,
  Muhlleitner:2014vsa}.  Finally, we proposed a way to extend to the
NMSSM the estimates of the theory uncertainty associated with
uncomputed higher-order effects that had previously been developed for
the MSSM.

We found that, in the NMSSM scenario with heavy BSM particles, the
matching condition for the quartic Higgs coupling splits neatly into a
part that is in common with the MSSM, and an NMSSM-specific part which
vanishes for $\lambda\rightarrow 0$. We studied the numerical impact
on the Higgs-mass prediction of the different contributions to the
NMSSM-specific part of the matching condition, and found that the
one-loop and two-loop-QCD contributions modify only moderately, and
only for quite large values of $\lambda$, the leading behavior driven
by the tree-level contribution. We stress that the smallness of these
effects is in fact a desirable feature of the EFT calculation of the
Higgs mass, in which the logarithmically enhanced corrections are
accounted for by the evolution of the parameters between the SUSY
scale and the EW scale, and high-precision calculations at the EW
scale can be borrowed from the SM.

Turning to the modern approach of treating the Higgs mass as an input
rather than an output of the calculation, we illustrated how the
requirement of a correct prediction for $M_h$ can be used to constrain
some of the yet-unmeasured parameters of the NMSSM. Focusing on the
$\tan\beta$--$\lambda$ plane, we noticed how the shape of the allowed
regions can change drastically depending on the choice of the
remaining parameters. This is a well-known aspect of the NMSSM, in
which the Higgs sector depends on a relatively large number of
parameters already at the tree level. However, a systematic
phenomenological study of the constraints that the Higgs-mass
prediction imposes on the parameter space of the NMSSM goes beyond the
scope of this paper. What we provide here is a set of fully analytic
formulas for the matching condition for $\lSM$, available on request
in electronic form. Our results can be used to implement the
resummation of the large logarithmic corrections in the existing
public codes for the Higgs-mass calculation in the
NMSSM~\cite{Ellwanger:2004xm, Ellwanger:2005dv,Staub:2008uz,
  Staub:2009bi, Staub:2010jh, Staub:2012pb, Staub:2013tta,
  Porod:2003um, Porod:2011nf, Baglio:2013iia, Allanach:2001kg,
  Allanach:2013kza, Athron:2014yba, Athron:2017fvs, Heinemeyer:1998yj,
  Hahn:2009zz, Bahl:2018qog}, bringing the accuracy of those codes
closer to what has already been attained for the MSSM.

\section*{Acknowledgments}

We thank M.~Gabelmann for useful communications concerning
ref.~\cite{Gabelmann:2018axh}.  The work of M.~G.~and P.~S.~is
supported in part by French state funds managed by the Agence
Nationale de la Recherche (ANR), in the context of the grant
``HiggsAutomator'' (ANR-15-CE31-0002). M.~G.~also acknowledges support
from the ANR grant ``DMwithLLPatLHC'' (ANR-21-CE31-0013).

\vfill
\newpage

%\begin{Appendix}
%\section*{Appendix}
%\end{Appendix}
%\vfill
%\newpage

\bibliographystyle{utphys}
\bibliography{BGS_rev}

\end{document}